\documentclass[a4paper,aps,pra,showpacs,superscriptaddress,twocolumn,nofootinbib]{revtex4-1}

\usepackage{graphicx,graphics,epsfig}   
\usepackage{dcolumn}    
\usepackage{bm}         
\usepackage{amsmath}    
\usepackage{verbatim}   
\usepackage{color}      
\usepackage{subfigure}  
\usepackage{times,natbib}
\usepackage{amsmath,amsfonts,amssymb,graphics,graphics,color,times}

\usepackage{latexsym}
\usepackage{amsmath}
\usepackage{amssymb}
\usepackage{amsfonts}
\usepackage{amsthm}
\usepackage{mathrsfs}
\usepackage{color,verbatim,graphics}
\usepackage{psfrag}
\DeclareMathAlphabet{\mathrsfs}{U}{rsfs}{m}{n}
\DeclareMathAlphabet{\mathpzc}{OT1}{pzc}{m}{it}
\DeclareMathAlphabet{\matheus}{U}{eus}{m}{n}
\DeclareMathAlphabet{\mathbbold}{U}{bbold}{m}{n}

\def\one{\leavevmode\hbox{\small1\normalsize\kern-.33em1}}

\newcommand{\ba}{\begin{eqnarray}}
\newcommand{\ea}{\end{eqnarray}}
\newcommand{\ban}{\begin{eqnarray*}}
\newcommand{\ean}{\end{eqnarray*}}

\begin{document}

\title{Closing the detection loophole in tripartite Bell tests using the W state}

\author{K\'aroly F. P\'al}
\affiliation{Institute for Nuclear Research, Hungarian Academy of Sciences, H-4001 Debrecen, P.O. Box 51, Hungary}

\author{Tam\'as V\'ertesi}
\affiliation{Institute for Nuclear Research, Hungarian Academy of Sciences, H-4001 Debrecen, P.O. Box 51, Hungary}
\affiliation{D\'epartement de Physique Th\'eorique, Universit\'e de Gen\`eve, 1211 Gen\`eve, Switzerland}
\date{\today}


\begin{abstract}
We study the problem of closing the detection loophole in
three-qubit Bell tests, the experimentally most relevant case
beyond the usual bipartite scenario, and show that the minimal
detection efficiencies required can be considerably lowered compared to the
two-qubit case. The lowest reported detection
efficiency thresholds for two and three qubits so far are $\sim66.7\%$ and $60\%$, respectively.
Using the three-qubit W state and a 3-setting Bell inequality, we beat these thresholds
and with an 8-setting Bell inequality we reach $50.13\%$. We also
investigate generic three-qubit states which allow us to attain a
detection efficiency of $50\%$ in a 4-setting Bell test. We
conjecture that the limit of $50\%$ is unbeatable using
three-qubit states and any number of measurements.
\end{abstract}

\maketitle

\section{Introduction}

One of the most surprising features of quantum mechanics is the
prediction that distant parties performing measurements on a shared
entangled state are able to generate correlations which rule out
any local hidden variables explanation. These nonlocal correlations can
be witnessed by the violation of Bell inequalities
\cite{Bell,review}. By now, many Bell experiments using various
matter systems have been performed (e.g., recently in \cite{rowe,ansmann,hofmann,christensen,giustina,carvacho})
providing strong indication for the existence of nonlocal
correlations in nature \cite{aspect}. However, imperfections in
the technical implementations of these experiments make it possible to
reproduce the experimental data by local hidden variables model. In order
to avoid such a classical explanation, all possible loopholes have to be closed
simultaneously in a Bell experiment. There are two main technical loopholes, the locality
loophole and the detection loophole. The former one can be closed if there is space-like separation between the
observers such that no signal can propagate from one observer to the other. This condition could only be met so far in photonic experiments \cite{adr,weihs,tittel}.

In the present paper, we would like to address the latter one, the so-called
detection loophole. This loophole is most relevant in Bell tests
which use photons, in which case measurements frequently give
undetected events. These no-click events have to be included in
the observed data, and nonlocal correlations are witnessed detection loophole-free only if there is no local
hidden variables model of the full statistics taking into account
the no-click events as well \cite{pearle}. The detection loophole has been closed
in different physical systems such as ions \cite{rowe}, superconductors \cite{ansmann}, atoms \cite{hofmann}, and more recently in photonic systems as well \cite{christensen,giustina}.

As we have seen, the only system where both primary loopholes have been closed are photons, albeit these were not closed in the same experiment. Though, important steps have been made both experimentally (see references above) and both theoretically \cite{theory}, such a loophole-free violation of a Bell inequality has not been performed yet. A comprehensive review on this subject can be found in Ref.~\cite{larsson_review}.

Let us mention that closing the detection loophole is also relevant from a practical point of view. The more recent development of device-independent quantum information protocols crucially rely on a detection loophole-free violation of Bell inequalities. In these protocols, there is no need to assume any knowledge regarding the
internal workings of the experimental devices used (see \cite{scarani} for a recent review of the field). For instance, it would allow two distant parties to establish a certified secret key \cite{key}, generate genuinely random numbers \cite{randomness}, or perform black-box state tomography \cite{tomography}.

In order to close the detection loophole, we construct Bell inequalities which are suited to
reveal nonlocality using detectors with low efficiencies. We will consider the relatively unexplored case of
three-party Bell inequalities involving finite detection efficiencies. In particular, we will
focus on the case when each party detects particles with the same
$\eta$ detection efficiency. The critical detection efficiency
$\eta_{crit}$, below which nonlocality cannot be guaranteed depends both
on the Bell inequality considered and the quantum state used in
the Bell test.

In the two-party case, $\eta_{crit}\sim 66.7\%$ \cite{eberhard} is required to violate the
Clauser-Horne-Shimony-Holt (CHSH) inequality \cite{chsh} with a partially
entangled two-qubit state. For two-qubits, to the best of our knowledge, there is
no known Bell inequality (with possibly more than two settings and
more than two outputs), which would give a lower threshold. Using
4-dimensional quantum states and a four-setting Bell inequality,
this threshold can be slightly lowered (down to $\sim 61.8\%$
\cite{VPB}), however, it is still too high when compared to
efficiencies achievable with current technology.

One possible approach to go below these threshold values is to
consider multipartite Bell tests, i.e., more than two observers.
Buhrman et al.~\cite{BHMR} and more recently
Ref.~\cite{PVB} have showed that an arbitrarily small efficiency
$\eta$ can be tolerated as the number of parties $n$ and the
number of settings $m$ become large. However, these results are
interesting mainly from a theoretical point of view. Indeed, in
the experimentally more relevant case of small number of settings,
the known results are less promising. For instance, if the number
of settings per party is fixed to two ($m=2$), the lowest
threshold efficiencies using the Mermin inequality~\cite{mermin}
and its generalized version~\cite{MABK} were shown to approach
$\eta_{crit}=50\%$ for large $n$ \cite{CRV}. The same limit can be
approached if we use the many-site generalization of the
Clauser-Horne inequality~\cite{LS}. Also, a multipartite
two-setting Bell test based on single-photon entanglement (i.e., a W-state shared between multiple parties) was
shown to approach $\eta_{crit}\simeq66.7\%$ for large $n$ \cite{CB}.
These above examples considered large number of parties and two
settings. There exist other constructions for class of multipartite
two-setting inequalities (e.g., \cite{hillery,zukowski}).
Note, however, that due to Ref.~\cite{massar} the
critical efficiency for two-setting inequalities cannot be lower than
$n/(2n-1)$. Hence, none of these inequalities may allow us to go below
$\eta_{crit}=60\%$ for three parties and below $\eta_{crit}=50\%$ for infinite
number of parties.

In contrast to two settings and large number of parties, the case of more than two settings per party and moderate number of parties is much less explored. Indeed, in the case
of three parties ($n=3$) and a few number of settings $m$, which is
the experimentally most interesting setup beyond the usual
two-party scenario, only a few results are known. To the best of
our knowledge, for three parties the lowest detection efficiency
is attained in Ref.~\cite{LS} giving $\eta_{crit}=60\%$ using $m=2$ settings.
The aim of this paper is to go beyond two measurement settings per
party, which opens the door to more efficient multisetting Bell
inequalities. In particular, we explore numerically the best
detection efficiencies for the emblematic three-qubit W state
\cite{dur}, and also perform detailed numerical search when the
underlying state is a more general symmetric 3-qubit pure state. Note
that the search for critical detection efficiencies using the
famous Greenberger-Horne-Zeilinger (GHZ) \cite{ghz} state was carried out recently in Ref.~\cite{PVB},
attaining the lowest efficiency $\eta_{crit}=12/17\simeq70.59\%$ so far using
$m=17$ settings per party (for an explicit construction of the Bell inequality, please
see the website~\cite{web}). Before this work the best bound of
$\eta_{crit} = 75\%$ for a GHZ state was provided by Larsson \cite{Larsson}
using the Mermin inequality.

Here we report a considerable improvement over the above values by
showing that detection efficiencies as low as $50\%$ can be
tolerated in tripartite Bell tests featuring a reasonable number
of measurements. However, our setups turn out to be very fragile
to noise, hence, we believe that the experimental implementation
remains a challenging issue.

\section{Setup}

We consider a Bell scenario with three observers ($n=3$), Alice,
Bob, and Cecil, who carry out experiments in distant laboratories.
Each observer can choose among $m$ possible inputs and receive two possible outcomes. Let
us identify the inputs of the three parties with
$i,j,k=1,\ldots,m$ which correspond to a set of $m$ possible
measurements $\{A_i\}$, $\{B_j\}$, $\{C_k\}$ for each party.
Without loss of generality, we can label with $+1$ and $-1$ the
two different outcomes $\alpha,\beta$, and $\gamma$ for the respective parties.
The experiment is fully characterized by the conditional
probabilities $P(\alpha\beta\gamma|A_iB_jC_k)$. We use the shorthand notation
$P(A_iB_jC_k)\equiv P(111|A_iB_jC_k)$ and similarly for a subset
of the parties, such as $P(A_iB_j)\equiv P(11|A_iB_j)$ and
$P(A_i)\equiv P(1|A_i)$, etc. It can be seen that these
probabilities fully determine the joint distribution
$P(\alpha\beta\gamma|A_iB_jC_k)$, hence it is enough to consider them.

Throughout this work we stick to symmetric Bell inequalities, that is, inequalities which are symmetric for all permutations of the parties. In addition, our Bell inequalities will not contain single party marginal terms, they are built up only by two-particle and three-particle correlation terms. We will also assume without loss of generality that the classical bound of the Bell inequalities are zero. The Bell inequalities considered in Refs.~\cite{LS},~\cite{PVB} are similarly restricted. As we will see, this simplification allows us to treat the problem with the tools of linear programming. We can write such a Bell inequality as:
\begin{align}
&\sum_{i,j=1}^{m}M^{(2)}_{ij}[P(A_iB_j)+P(A_iC_j)+P(B_iC_j)]\nonumber\\
&+\sum_{i,j,k=1}^{m}M^{(3)}_{ijk}P(A_iB_jC_k)\leq 0,
\label{eq:Bellineq}
\end{align}
where
\begin{align}
M^{(3)}_{ijk}&=M^{(3)}_{ikj}=M^{(3)}_{jik}=M^{(3)}_{jki}=M^{(3)}_{kij}=M^{(3)}_{kji}\nonumber\\
M^{(2)}_{ij}&=M^{(2)}_{ji}, \label{eq:coeffsymm}
\end{align}
and the Bell coefficients $M^{(3)}_{ijk}$ and $M^{(2)}_{ij}$ are
chosen such that the classical bound is zero.

\subsection{Local bound}

Let us first compute the local limit of the above Bell inequality~(\ref{eq:Bellineq}) allowing any classical mechanism. In order to do that, it is enough to consider deterministic strategies: Each of the parameters $a_i$,
$b_i$ and $c_i$, where $i$ runs from 1 to $m$, may take the value of either $0$ or $1$, and a
deterministic strategy is defined by a particular choice. This
corresponds to a definite outcome for each measurement value for
each party. For example, $a_i=1$ means that the probability for
Alice to get the value $+1$ for her $i$th measurement is one, that is
$P(A_i)=1$.

To set the classical bound of the Bell inequality~(\ref{eq:Bellineq}) to zero, we must ensure that
\begin{equation}
\sum_{i,j=1}^{m}M^{(2)}_{ij}(a_ib_j+a_ic_j+b_ic_j)+
\sum_{i,j,k=1}^{m}M^{(3)}_{ijk}a_ib_jc_k\leq 0 \label{eq:detstrat}
\end{equation}
for all deterministic strategies. Eq.~(\ref{eq:detstrat}) gives $2^{3m}$ linear
constraints for the Bell coefficients. Due to the permutational
symmetry in Eq.~(\ref{eq:coeffsymm}), two strategies which may be
derived from each other by swapping the strategies of any two
participants (e.g. by swapping the values of $a_i$ and $b_i$)
lead to the same constraint, which makes it possible to reduce the
number of constraints. Also, Eq.~(\ref{eq:detstrat}) is trivially
fulfilled for any strategy assigning nonzero values for only one
of the participants. We note that it follows from
Eq.~(\ref{eq:detstrat}) that $M^{(2)}_{ij}\leq 0$. We may get this
from strategy $a_i=1$, $b_j=1$, while all other $a$ and $b$, and
all $c$ values are zero.

\subsection{Quantum bound}

Now let us consider the quantum case. The maximum quantum violation of a two-outcome Bell inequality (i.e. the one presented in (\ref{eq:Bellineq})) is always attained by von Neumann measurements~\cite{cleve}. Moreover, it is sufficient to restrict ourselves to pure states $|\psi\rangle$, that is
$\hat\rho=|\psi\rangle\langle\psi|$. Then
\begin{align}
P(A_iB_j)&=\langle \psi|\hat A_i\otimes \hat B_j\otimes \hat I|\psi\rangle\nonumber\\
P(A_iC_j)&=\langle \psi|\hat A_i\otimes \hat I\otimes \hat C_j|\psi\rangle\nonumber\\
P(B_iC_j)&=\langle \psi|\hat I\otimes \hat B_i\otimes \hat C_j|\psi\rangle\nonumber\\
P(A_iB_jC_k)&=\langle \psi|\hat A_i\otimes \hat B_j\otimes \hat
C_k|\psi\rangle, \label{eq:Qprobab}
\end{align}
where $\hat A_i$, $\hat B_j$ and $\hat C_k$ are the measurement
operators of Alice, Bob and Cecil, respectively, projecting onto
the subspace corresponding to outcome $+1$ in the subspace of the
participant concerned, and $\hat I$ is the unity operator in that
subspace. Along this study, we will restrict ourselves to 3-qubit states,
hence the measurement operators $\hat A_i$, $\hat B_j$ and $\hat C_k$ are in fact projectors in the qubit space.

\subsection{Quantum case with limited detection efficiency}

Let us consider the quantum case when all participants detect
their particles with the same limited detection efficiency $\eta$.
As a side remark, we note that interesting results have been obtained in the asymmetric case, that is, when the parties feature different efficiencies~\cite{asym} or when measurements corresponding to the same party have different efficiencies \cite{garbarino}.
In our symmetric scenario, the participants agree to output $-1$ in case of no detection. In this case, we get
the joint probabilities of detecting outcome $+1$ by two and by all
three participants if we multiply the probabilities of
Eq.~(\ref{eq:Qprobab}) by $\eta^2$ and by $\eta^3$, respectively,
that is by the probability of the detection of the particles
concerned. Then the condition for the violation of the Bell
inequality in Eq.~(\ref{eq:Bellineq}) can be written as:
\begin{equation}
\langle\psi|\hat{\cal M}_\eta|\psi\rangle\equiv\eta^2{\cal
M}^{(2)}+\eta^3{\cal M}^{(3)}>0, \label{eq:violate}
\end{equation}
where
\begin{align}
{\cal M}^{(2)}\equiv&\sum_{i,j=1}^{m}M^{(2)}_{ij}\big
(\langle\psi|\hat A_i\otimes\hat B_j\otimes\hat I
|\psi\rangle\nonumber\\
&+\langle\psi|\hat A_i\otimes \hat I\otimes \hat C_j|\psi\rangle+
\langle\psi|\hat I\otimes \hat B_j\otimes \hat C_k|\psi\rangle\big )\label{eq:2partcontr}\\
{\cal M}^{(3)}\equiv&\sum_{i,j,k=1}^{m}M^{(3)}_{ijk}
\langle\psi|\hat A_i\otimes\hat B_j\otimes\hat C_k|\psi\rangle,
\label{eq:3partcontr}
\end{align}
and $\hat{\cal M}_\eta$ is the effective Bell operator at $\eta$
efficiency. As we have shown earlier, $M^{(2)}_{ij}\leq 0$,
therefore ${\cal M}^{(2)}\leq 0$. Therefore, if $\eta$ is very
small, according to Eq.~(\ref{eq:violate}), there is no Bell violation.
The critical detector efficiency, above which the violation may be
detected is:
\begin{equation}
\eta_{crit}=-\frac{{\cal M}^{(2)}}{{\cal M}^{(3)}}.
\label{eq:etacrit}
\end{equation}

To find the Bell inequality which minimizes $\eta_{crit}$ in case
of a particular choice of the state and the measurement operators
is a problem of standard linear programming. To ensure that the
classical bound is zero, the set of linear constraints given by
Eq.~(\ref{eq:Bellineq}) must be satisfied. As the Bell
coefficients may be multiplied by any positive number, we may fix
the norm by fixing the value of ${\cal M}^{(2)}$.  We may choose
any negative number. In particular, let us choose ${\cal
M}^{(2)}=-1$. This provides an additional linear constraint. Then
we must maximize ${\cal M}^{(3)}$, which is a linear expression
for the Bell coefficients. The symmetries according to
Eq.~(\ref{eq:coeffsymm}) are further linear constraints to be
enforced, but instead of doing that, we may restrict ourselves to
coefficients $M^{(3)}_{ijk}$, with $i\leq j\leq k$ and $
M^{(2)}_{ij}$, with $i\leq j$, and rewrite the constraints and the
expression to be maximized in terms of these independent
parameters. This way we get a much smaller problem to solve.

Let the set of measurement operators be the same for all parties,
and let us confine ourselves to real measurement operators. This particular restriction was also proved to be useful in other studies for exploring nonlocality of the W state \cite{Wpapers}.
In this case the operator $\hat A_i=\hat B_i=\hat C_i$ can be
characterized by a single real variable $\Phi_i$:
\begin{align}
&\hat
A_i|0\rangle=\frac{1}{2}(1-\cos\Phi_i)|0\rangle-\frac{1}{2}\sin\Phi_i|1\rangle
\equiv c_i^-|0\rangle+s_i|1\rangle\nonumber\\
&\hat
A_i|1\rangle=-\frac{1}{2}\sin\Phi_i|0\rangle+\frac{1}{2}(1+\cos\Phi_i)|1\rangle
\equiv s_i|0\rangle+c_i^+|1\rangle. \label{eq:measops}
\end{align}
If $\Phi_i=0$, the measurement gives value $+1$ with probability one
for the $|1\rangle$ state.

Let the quantum state be also symmetric in terms of the permutations of the parties. One such a state is the 3-qubit
GHZ state~\cite{ghz}, which case has been already investigated thoroughly \cite{Larsson,PVB}. In this paper our primary concern is the 3-qubit W state \cite{dur} but we also study generic symmetric 3-qubit states. In the following section we focus on the W state (Sec.~\ref{Wstate}) and then we move on to investigate the more general case in Sec.~\ref{Wlikestate}. Our main results concerning the found detection efficiency thresholds are summarized in Table~I and Table~II for the W state and the generic 3-qubit states, respectively.

\section{Detection efficiencies using the W state}\label{Wstate}
The W state is defined by \cite{dur}:
\begin{equation}
|W\rangle=\frac{1}{\sqrt{3}}(|001\rangle+|010\rangle+|100\rangle),
\label{eq:wstate}
\end{equation}
where we have used the shorthand notation:
\begin{equation}
|\alpha\beta\gamma\rangle\equiv|\alpha\rangle\otimes|\beta\rangle\otimes|\gamma\rangle.
\label{eq:tensprodnotat}
\end{equation}
Now, by using
Eqs.~(\ref{eq:measops},\ref{eq:wstate},\ref{eq:tensprodnotat}), it
is straightforward to calculate the quantum conditional
probabilities appearing in
Eqs.~(\ref{eq:2partcontr},\ref{eq:3partcontr}):
\begin{align}
&\langle W|\hat A_i\otimes\hat A_j\otimes\hat I|W\rangle=
\frac{1}{3}(2s_is_j+c_i^-c_j^++c_i^+c_j^-+c_i^-c_j^-)\nonumber\\
&\langle W|\hat A_i\otimes\hat A_j\otimes\hat A_k|W\rangle=
\frac{2}{3}(c_i^-s_js_k+s_ic_j^-s_k+s_is_jc_k^-)\nonumber\\
&+\frac{1}{3}(c_i^-c_j^-c_k^++c_i^-c_j^+c_k^-+c_i^+c_j^-c_k^+).
\label{eq:wmatelems}
\end{align}
If the number of measurement settings per party is small, we can
scan the space of measurement angles with an even step size, and
solve the linear programming problem for each set of angles. In
each case the optimal Bell inequality we arrive at has to be a
tight one in the symmetrized probability space. We refer to Ref.~\cite{sym}
for the framework of symmetric Bell inequalities and to further studies which makes
use of this framework \cite{symfurther} reducing considerably the complexity of the problem.
There is a finite number of such inequalities, so we get the same
solution for a whole range of angles. Therefore, if our step size
is not too large, we will certainly get the Bell inequality that
gives the smallest critical efficiency with the $|W\rangle$ state.
Then for the known inequality we may calculate the optimum
measurement angles. Also, due to the tightness, the Bell
coefficients can always be normalized such that they are integer
numbers.

Next we list our results for different number of settings, where the numerical study was carried out up to 8 settings per party.

\subsection{W state, $m=2$}\label{m2}

For two measurement settings per party we got the following Bell coefficients:
\begin{align}
M_{11}^{(2)}=-1 && M_{111}^{(3)}=2 && M_{112}^{(3)}=1 &&
M_{122}^{(3)}=-1. \label{eq:B222w}
\end{align}
Here we only show the values of the independent Bell coefficients,
that is $M^{(2)}_{ij}$ with $i\leq j$ and $M^{(3)}_{ijk}$, with
$i\leq j\leq k$. The values of the coefficients that can not be
derived from the coefficients given above by some permutation of
the parties (e.~g.\ $M_{12}^{(2)}$) are zero. This inequality is equivalent to the
inequality 22 in the list of Sliwa~\cite{Sliwa}.
The optimum angles for this inequality are $\Phi_1=2.28059$ and $\Phi_2=0.33432$, and
the critical efficiency is $\eta_{crit}=0.83747$. We will see
later that the $|W\rangle$ state is not the best choice for this
inequality.

\subsection{W state, $m=3,4,5$}\label{m345}

For $m=3$ and $m=4$ we have got the inequalities with the smallest $\eta_{crit}$
if we have chosen the measurement angles small. In the case of
$m=3$, the nonzero independent Bell coefficients of this
inequality are:
\begin{align}
&M_{11}^{(2)}=-6 && M_{23}^{(2)}=-3 && M_{123}^{(3)}=3 && M_{223}^{(3)}=2 \nonumber\\
&M_{233}^{(3)}=2, \label{eq:B333w}
\end{align}
while for $m=4$ we have got:
\begin{align}
&M_{12}^{(2)}=-6 && M_{34}^{(2)}=-2 && M_{112}^{(3)}=6 && M_{114}^{(3)}=-6 \nonumber\\
&M_{122}^{(3)}=6 && M_{123}^{(3)}=3 && M_{124}^{(3)}=3 && M_{134}^{(3)}=-1 \nonumber\\
&M_{223}^{(3)}=-6 && M_{234}^{(3)}=-1 && M_{334}^{(3)}=2 &&
M_{344}^{(3)}=2, \label{eq:B444w}
\end{align}
For both inequalities the optimal angles for all measurement
settings approach zero near the threshold efficiency. This
observation allows us to make some analytical considerations.

Let $x$ be small, and let us consider the measurement angles
proportional to this small number, that is $\Phi_i\equiv\phi_ix$.
Then, Eq.~(\ref{eq:wmatelems}) may be approximated as:
\begin{align}
&\langle W|\hat A_i\otimes\hat A_j\otimes\hat I|W\rangle\approx
\frac{1}{6}[1-\cos x(\phi_i+\phi_j)]+\frac{x^4}{48}\phi_i^2\phi_j^2\label{eq:wmatelemssmall1}\\
&\langle W|\hat A_i\otimes\hat A_j\otimes\hat A_k|W\rangle\approx
\frac{x^4}{48}(\phi_i\phi_j+\phi_i\phi_k+\phi_j\phi_k)^2.
\label{eq:wmatelemssmall2}
\end{align}
We have neglected terms sixth and higher order in $x$. We have
used Eq.~(\ref{eq:measops}) defining the quantities appearing in
Eq.~(\ref{eq:wmatelems}), which may be approximated at leading
order as $s_i\approx-\phi_i x/2$, $c_i^+\approx 1$ and
$c_i^-\approx\phi^2x^2/4$. Also, it is easy to see that
$2s_is_j+c_i^-c_j^++c_i^+c_j^-=[1-\cos(\Phi_i+\Phi_j)]/2$. Due to
Eq.~(\ref{eq:wmatelemssmall2}), ${\cal M}^{(3)}$ (see
Eq.~(\ref{eq:3partcontr})) is fourth order in $x$. Then, according
to Eq.~(\ref{eq:etacrit}), we may only get a finite value for
$\eta_{crit}$, if ${\cal M}^{(2)}$ defined in
Eq.~(\ref{eq:2partcontr}) is also fourth order in $x$. This is
true if whenever the $M_{ij}^{(2)}$ Bell coefficient is not zero,
the corresponding measurement angles satisfy $\phi_i+\phi_j=0$.

We may get the Bell inequalities of
Eqs.~(\ref{eq:B333w},\ref{eq:B444w}) by solving the linear
programming problem using the small angles limit, that is
Eqs.~(\ref{eq:wmatelemssmall1},\ref{eq:wmatelemssmall2}), when
calculating ${\cal M}^{(2)}$ and ${\cal M}^{(3)}$, and dropping
the overall factor $x^4$. In case of $m=3$ (Eq.~\ref{eq:B333w}),
we take $\phi_1=0$ and $\phi_2=-\phi_3=1$ (that is we choose
$x=\Phi_2)$. This way, there are no free parameters left. With
this choice $M_{11}^{(2)}$ and $M_{23}^{(2)}$ may take a nonzero
value, as $\Phi_1=-\Phi_1=0$ and $\Phi_2=-\Phi_3=x$. Indeed, these
are the nonzero $M_{ij}^{(2)}$ coefficients in
Eq.~(\ref{eq:B333w}). Actually, the solution of the linear
programming problem in this case is not unique, there are other
Bell inequalities leading to the same $\eta_{crit}$. We have shown
the one having the smallest number of nonzero Bell coefficients.
Now, from Eq.~(\ref{eq:etacrit}) we can easily calculate the value
of $\eta_{crit}$. Using the measurement angles defined above,
$\langle W|\hat A_1\otimes\hat A_1\otimes\hat I|W\rangle\approx 0$
and $\langle W|\hat A_2\otimes\hat A_3\otimes\hat
I|W\rangle\approx x^4/48$ (see Eq.~\ref{eq:wmatelemssmall1}).
Furthermore, $\langle W|\hat A_1\otimes\hat A_2\otimes\hat
A_3|W\rangle\approx \langle W|\hat A_2\otimes\hat A_2\otimes\hat
A_3|W\rangle= \langle W|\hat A_2\otimes\hat A_3\otimes\hat
A_3|W\rangle\approx x^4/48$. Also, due to the permutational
symmetry of the state $|W\rangle$, the matrix elements are the
same for all permutations of the operators. Therefore, by
substituting the values for the measurement angles and the Bell
coefficients into Eq.~(\ref{eq:2partcontr}), we get ${\cal
M}^{(2)}=-18x^3/48$. Similarly, from Eq.~(\ref{eq:3partcontr}), we
arrive at ${\cal M}^{(3)}=30x^3/48$. Therefore,
$\eta_{crit}=3/5=0.6$. We have noted that this is not the only
Bell inequality with the same threshold efficiency. The reason is
that the quantum value does not depend on $M_{11}^{(2)}$,
$M_{111}^{(3)}$, as the matrix elements they are multiplied with
are zero being $\Phi_1=0$. The requirement of zero classical value
does not define uniquely these coefficients.

In the case of $m=4$, similarly to Eq.~(\ref{eq:B333w}) for $m=3$, Eq.~(\ref{eq:B444w})
can also be derived by using the small angles limit. Now, we
choose the measurement angles $\Phi_1=-\Phi_2=x$ and
$\Phi_3=-\Phi_4=\lambda x$. Now we have a single parameter
$\lambda$. It is enough to consider $|\lambda|\leq 1$. We get the
required inequality if we choose any value for $\lambda$ between
0.21 and 0.78. We show in the Appendix that the optimum is
$\lambda=0.466715$, which is a root of a fifth order equation, and
then $\eta_{crit}=0.509036$.

We have also derived the optimal $m=5$ Bell inequality similarly to the smaller
ones in section~\ref{m345}, with measurement angles $\Phi_1=0$,
$\Phi_2=-\Phi_3=x$ and $\Phi_4=-\Phi_5=\lambda x$. It turned out
to be equivalent to the $m=4$ case, so we got no improvement on
the critical efficiency.

\subsection{W state, $m_A=3$ and $m_B=m_C=2$}\label{masym}

If we do not require permutational symmetry, we may create a Bell inequality with the same
$\eta_{crit}=0.6$ as for the $m_A=m_B=m_C=3$ case using the $|W\rangle$ state with only two
measurement settings for Bob and Cecil. As before, Alice's
measurement settings $\hat A_1$, $\hat A_2$ and $\hat A_3$ are
characterized by $\Phi^A_1=0$, $\Phi^A_2=x$ and $\Phi^A_3=-x$,
respectively. However, for Bob and Cecil the measurement angles
will be chosen as $\Phi^B_1=0$, $\Phi^B_2=x$ and $\Phi^C_1=0$,
$\Phi^C_2=-x$, respectively. Here we used the upper indices to
distinguish between the parties. The asymmetric inequality will
have the same quantum value as the symmetric one for any $\eta$,
if the sum of the Bell coefficients multiplying matrix elements
that have the same numerical value are the same for both
inequalities. At the same time we must ensure that the classical
bound is also the same, that is zero. In the case of a known
symmetric inequality, these requirements define a set of linear
constraints for the coefficients of the asymmetric one. It is a
problem of linear programming to decide whether these constraints
can be satisfied or not. In the present case the problem is
solvable, the simplest Bell inequality we have got, after dividing
each coefficient by a factor of six is:
\begin{align}
S\equiv-&P(11|A_1B_1)-P(11|A_1C_1)-P(11|B_1C_1)\nonumber\\
-&P(11|A_3B_2)-P(11|A_2C_2)-P(11|B_2C_2)\nonumber\\
+&P(111|A_1B_2C_2)+P(111|A_2B_1C_2)+P(111|A_3B_2C_1)\nonumber\\
+&P(111|A_2B_2C_2)+P(111|A_3B_2C_2)\leq 0. \label{eq:Bell322}
\end{align}

We have also tried to derive asymmetric Bell inequalities with a
smaller number of measurement settings for some of the parties
from the $m=4$ case given by Eq.~(\ref{eq:B444w}), and also from
the inequalities we will show later, but we have found no solution
for the problem involved.

\subsection{W state, $m\ge 6$}\label{m6}
For $m=6$, using $\Phi_1=-\Phi_2=x$, $\Phi_3=-\Phi_4=\mu
x$ and $\Phi_5=-\Phi_6=\nu x$, we got a new inequality, with
$\eta_{crit}=0.502417$, marginally better than before. Now the
optimal choice for the parameters is $\mu=0.495815$ and
$\nu=0.295435$ (see Appendix). The nonzero independent Bell
coefficients of this $m=6$ inequality are:
\begin{align}
&M_{12}^{(2)}=-18 && M_{34}^{(2)}=-18 && M_{56}^{(2)}=-18 && M_{112}^{(3)}=18 \nonumber\\
&M_{114}^{(3)}=-18 && M_{122}^{(3)}=18 && M_{123}^{(3)}=9 && M_{124}^{(3)}=9 \nonumber\\
&M_{136}^{(3)}=-9 && M_{156}^{(3)}=8 && M_{223}^{(3)}=-18 && M_{245}^{(3)}=-9 \nonumber\\
&M_{256}^{(3)}=4 && M_{334}^{(3)}=18 && M_{336}^{(3)}=-18 && M_{344}^{(3)}=18 \nonumber\\
&M_{345}^{(3)}=9 && M_{346}^{(3)}=9 && M_{356}^{(3)}=1 && M_{445}^{(3)}=-18 \nonumber\\
&M_{456}^{(3)}=5 && M_{556}^{(3)}=4 && M_{566}^{(3)}=8,
\label{eq:B666w}
\end{align}

For $m=7$ we have got no further improvement.

For $m=8$ there are three free parameters. The angles are given as $\Phi_1=-\Phi_2=x$,
$\Phi_3=-\Phi_4=\rho x$, $\Phi_5=-\Phi_6=\sigma x$ and $\Phi_7=-\Phi_8=\tau x$. The
optimal choice of the parameters is $\rho=0.498442$, $\sigma=0.306395$ and
$\tau=0.169989$. Then $\eta_{crit}=0.501338$. The nonzero independent coefficients are:
\begin{align}
&M_{12}^{(2)}=-6 && M_{34}^{(2)}=-6 && M_{56}^{(2)}=-6 && M_{78}^{(2)}=-6 \nonumber\\
&M_{112}^{(3)}=6 && M_{114}^{(3)}=-6 && M_{122}^{(3)}=6 && M_{123}^{(3)}=3 \nonumber\\
&M_{124}^{(3)}=3 && M_{136}^{(3)}=-3 && M_{223}^{(3)}=-6 && M_{245}^{(3)}=-3 \nonumber\\
&M_{334}^{(3)}=6 && M_{336}^{(3)}=-6 && M_{344}^{(3)}=6 && M_{345}^{(3)}=3 \nonumber\\
&M_{346}^{(3)}=3 && M_{358}^{(3)}=-3 && M_{378}^{(3)}=2 && M_{445}^{(3)}=-6 \nonumber\\
&M_{467}^{(3)}=-3 && M_{478}^{(3)}=2 && M_{556}^{(3)}=6 && M_{558}^{(3)}=-6 \nonumber\\
&M_{566}^{(3)}=6 && M_{567}^{(3)}=3 && M_{568}^{(3)}=3 && M_{578}^{(3)}=1 \nonumber\\
&M_{667}^{(3)}=-6 && M_{678}^{(3)}=1 && M_{778}^{(3)}=2 && M_{788}^{(3)}=2,
\label{eq:B888w}
\end{align}

We have not tried any larger numbers of settings, the number of constraints
are too large. We may have got further improvement,
but we do not expect we could go below 0.5 with the critical
efficiency.

We summarized critical detection efficiencies we found in this paper for the 3-qubit W state in Table~\ref{table:bestdetW}.

\begin{table}[t]
\begin{tabular}{c|c c}
\hline
$\text{settings}$&$\eta_{crit}$&$\text{equation}$\\
\hline
222&0.83747&(\ref{eq:B222w})\\
223&0.6&(\ref{eq:Bell322})\\
333&0.6&(\ref{eq:B333w})\\
444&0.509036&(\ref{eq:B444w})\\
666&0.502417&(\ref{eq:B666w})\\
888&0.501338&(\ref{eq:B666w})\\
\hline
\end{tabular}
\caption{Table for critical detection efficiencies using the W state. The numbers in
brackets refer to the Bell inequalities. In the first
column $ijk$ refers to the respective number of settings $i$, $j$, and $k$ for the
parties Alice, Bob, and Cecil. Detection efficiency thresholds are rounded up
to five digits.} \vskip 0.2truecm
\label{table:bestdetW}
\end{table}

\section{Detection efficiencies for symmetric 3-qubit states}
\label{Wlikestate}

By considering a more general symmetric state we have been able to reach $\eta_{crit}=0.5$ exactly already with $m=4$. But we found improvement even for $m=3$.
The state considered is:
\begin{equation}
|\psi\rangle=\cos\alpha|W\rangle+\sin\alpha|111\rangle.
\label{eq:psistate}
\end{equation}
This state is also symmetric for the permutations of the parties,
therefore, the matrix elements of the tensor products of single
party operators will not depend on the order of those operators.
In the above state we find that the weight of
$|111\rangle$ goes to zero as the threshold efficiency is
approached. Like before, the measurement angles also vanish at
$\eta_{crit}$.

Now, besides the matrix elements calculated with the $|W\rangle$
state (see Eq.~(\ref{eq:wmatelems}), and
Eqs.~(\ref{eq:wmatelemssmall1},\ref{eq:wmatelemssmall2})) for the
conditional probabilities of Eq.~(\ref{eq:Qprobab}) appearing in
Eqs.~(\ref{eq:2partcontr},\ref{eq:3partcontr}) we also need:
\begin{align}
&\langle W|\hat A_i\otimes\hat A_j\otimes\hat I|111\rangle=
\frac{1}{\sqrt{3}}s_is_j\approx\frac{x^2}{4\sqrt{3}}\phi_i\phi_j\nonumber\\
&\langle W|\hat A_i\otimes\hat A_j\otimes\hat A_k|111\rangle=
\frac{x^2}{4\sqrt{3}}(c_i^+s_js_k+s_ic_j^+s_k+s_is_jc_k^+)\nonumber\\
&\approx\frac{x^2}{4\sqrt{3}}(\phi_i\phi_j+\phi_i\phi_k+\phi_j\phi_k)\nonumber\\
&\langle 111|\hat A_i\otimes\hat A_j\otimes\hat I|111\rangle=c_i^+c_j^+\approx 1\nonumber\\
&\langle 111|\hat A_i\otimes\hat A_j\otimes\hat
A_k|111\rangle=c_i^+c_j^+c_k^+\approx 1. \label{eq:othermatelems}
\end{align}
These matrix elements, including their limits for small angles,
may be calculated similarly to the ones given in
Eqs.~(\ref{eq:wmatelems},\ref{eq:wmatelemssmall1},\ref{eq:wmatelemssmall2}).
We have also used the same notations. For small angles we have
kept only the leading order terms. Eq.~(\ref{eq:othermatelems})
shows that the matrix elements $\langle W|\hat{\cal
M}_\eta|111\rangle$ and $\langle 111|\hat{\cal M}_\eta|111\rangle$
of the effective Bell operator (see Eq.~(\ref{eq:violate})) for
small measurement angles, that is for small $x$, will be second
and zeroth order in $x$, respectively. With $|\psi\rangle$ given
in Eq.~(\ref{eq:psistate}) we can write:
\begin{align}
&\langle\psi|\hat{\cal M}_\eta|\psi\rangle=\cos^2\alpha\langle
W|\hat{\cal M}_\eta|W\rangle\nonumber\\
&+2\sin\alpha\cos\alpha\langle W|\hat{\cal M}_\eta|111\rangle+\sin^2\alpha\langle 111|\hat{\cal M}_\eta|111\rangle.
\label{eq:effbell}
\end{align}
Let us make the same restriction as before, namely let
$M_{ij}^{(2)}=0$, whenever the corresponding measurement angles do
not satisfy $\phi_1+\phi_2=1$, which makes sure that $\langle
w|\hat{\cal M}_\eta|W\rangle$ fourth order in $x$. Then
$\langle\psi|\hat{\cal M}_\eta|\psi\rangle$ is also fourth order,
if the mixing angle $\alpha$ is taken proportional with $x^2$,
that is $\alpha=ax^2$. For small $x$ we may write:
\begin{align}
\langle\psi|\hat{\cal M}_\eta|\psi\rangle&\approx\langle
W|\hat{\cal M}_\eta|W\rangle+
2x^2\langle W|\hat{\cal M}_\eta|111\rangle a\nonumber\\
&+x^4\langle 111|\hat{\cal M}_\eta|111\rangle a^2.
\label{eq:effbellsmall}
\end{align}
With this choice, all matrix elements appearing in ${\cal
M}^{(2)}$ and ${\cal M}^{(3)}$ according to
Eqs.~(\ref{eq:2partcontr},\ref{eq:3partcontr}) are fourth order in
$x$, and we may derive the Bell inequalities with the smallest
critical efficiency using linear programming exactly the same way
as we have done with the $|W\rangle$ state. There is one extra
parameter $a$ characterizing the mixing angle. From
Eq.~(\ref{eq:effbellsmall}) it is easy to determine the optimum
choice for this parameter. The equation defines a parabola as a function of $a$, and its maximum value is given as
\begin{equation}
\label{param_a}
a=-\langle W|\hat{\cal M}_\eta|111\rangle/x^2\langle 111|\hat{\cal M}_\eta|111\rangle.
\end{equation}
 We note that $\langle 111|\hat{\cal M}_\eta|111\rangle\leq 0$ for small $x$, which
follows from the condition that the classical bound is zero, and
that all values of matrix elements involved are approximately one
(see Eq.~(\ref{eq:othermatelems})). Then the quantum value with
the optimum $a$ may be written as:
\begin{align}
\langle\psi|\hat{\cal M}_\eta|\psi\rangle&\approx\langle
W|\hat{\cal M}_\eta|W\rangle- \frac{\langle W|\hat{\cal
M}_\eta|111\rangle}{\langle 111|\hat{\cal M}_\eta|111\rangle}.
\label{eq:qsmallxlimit}
\end{align}

The optimum value of $a$ depends on the Bell coefficients to be
determined, so what we can do is to try some initial values for
$a$, determine the Bell inequality with linear programming,
calculate the optimum $a$ for this inequality, then repeat these
steps until convergency, which typically means just a few
iterations.

Let us first start with the smallest number of settings considered:

\subsection{Symmetric state, $m=2$}
Choosing the parameter $a$ according to (\ref{param_a}) in the state $|\psi\rangle$ in Eq.~(\ref{eq:psistate}), noting that $\alpha = ax^2$, we may get $\eta_{crit}=0.6$ in the limit of small measurement angles with
$m=2$ measurement settings per party. If we choose $\Phi_1=0$ and
$\Phi_2=x$, we get the same Bell inequality as we got with the
$|W\rangle$ state, we have shown in Eq.~(\ref{eq:B222w}). The
marginally small admixture of the $|111\rangle$ state lowered the value of
$\eta_{crit}$ from $0.83747$ to $0.6$, with considerably different
measurement angles. The inequality is the same as the three party one given
by Larsson \emph{et al.}~\cite{LS}, and which is number 22 on the
list of Sliwa~\cite{Sliwa}.
However, in \cite{LS} the state they
considered is the $|000\rangle$ state with a very small admixture of the $|W\rangle$ state, that is
their state approaches a separable state at the threshold efficiency. Also, in their case,
the second measurement angle is zero, and not the first one. Surprisingly, their very different solution
does lead to the same $\eta_{crit}=0.6$. We have calculated the
maximum violation of the inequality numerically for several detector efficiencies above $\eta_{crit}$.
It turned out that it is always enough to consider permutationally symmetric real states and
to take the same real measurement operators for each party. Therefore, the state can be written as
a linear combination of $|W\rangle$, $|111\rangle$ and $|000\rangle$ (the fourth independent
real symmetric state can always be eliminated by an appropriate choice of the local coordinates).

\begin{figure}[t]
\vspace{1.7cm}
\includegraphics[angle=-90, width=\columnwidth]{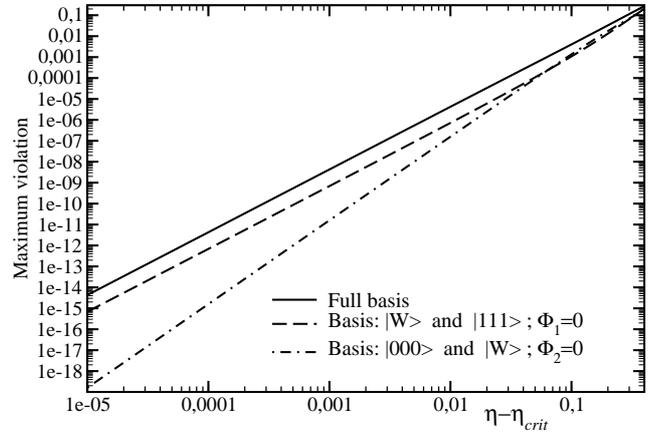} \caption{The maximum violation as a function of the detector efficiency for the Bell inequality with two settings per party.}
\label{fig:viol222}
\end{figure}

The maximum violation as a function of the detector efficiency is shown in Fig.~(\ref{fig:viol222}).
Near the threshold efficiency the maximum violation scales as the third power of
$\Delta\eta=\eta-\eta_{crit}$. The optimum state approaches the $|W\rangle$ state, while
the coefficients of the $|111\rangle$ and the $|000\rangle$ states are proportional to
$\Delta\eta$ and $\Delta\eta^{3/2}$, respectively. If we take the coefficient of the
$|000\rangle$ state exactly zero, the maximum violation remains basically the same.
Near the threshold the difference is negligible, and it is just a little more than $3\%$
around $\eta=0.9$. Therefore, the optimum solution may be reproduced almost exactly with
the state we have considered in the present paper. Near $\eta_{crit}$ the measurement angles
$\Phi_1$ and $\Phi_2$ scale as $\Delta\eta^{3/2}$ and $\Delta\eta^{1/2}$, respectively.
It is the first angle that tends to zero faster. If we take this angle exactly zero, as we
have done in this paper, the scaling behaviour of the maximum violation will not change,
but its value will be smaller by a factor approaching 6.25 near $\eta_{crit}$, and by a
factor of 1.33 at $\eta=1$ (see Fig.~(\ref{fig:viol222})). If we take the basis used at \cite{LS},
given by the $|000\rangle$ and the $|W\rangle$ states, the threshold efficiency
remains $0.6$, but near $\eta_{crit}$ we get much smaller violations: it will scale
as the fourth power of $\Delta\eta$. This time $\Phi_2$ goes to zero faster than
$\Phi_1$. If we take $\Phi_2=0$, it will hardly affect the violation near  $\eta_{crit}$,
while it will reduce it by about $30\%$ at $\eta=1$. The result is shown in Fig.~(\ref{fig:viol222})
We may conclude that for this inequality our solution is much closer to the optimal
arrangement than the one of Larsson \emph{et al.}~\cite{LS}. However, their approach may
directly be generalized to a larger number of parties.

\subsection{Symmetric state, $m=3$}
The independent Bell coefficients we got for $m=3$ are:
\begin{align}
&M_{11}^{(2)}=-2 && M_{23}^{(2)}=-1 && M_{111}^{(3)}=4 && M_{112}^{(3)}=1 \nonumber\\
&M_{113}^{(3)}=1 && M_{122}^{(3)}=-2 && M_{123}^{(3)}=1 && M_{133}^{(3)}=-2 \nonumber\\
&M_{223}^{(3)}=1 && M_{233}^{(3)}=1, \label{eq:B333wp111}
\end{align}
For this Bell inequality $\eta_{crit}=(19+\sqrt{937})/96\approx
0.516776$ (see Appendix), significantly smaller than the 0.6 value
we got with the $|W\rangle$ state for $m=3$.

\subsection{Symmetric state, $m\ge 4$}
For $m=4$ the coefficients are:
\begin{align}
&M_{12}^{(2)}=-2 && M_{34}^{(2)}=-2 && M_{112}^{(3)}=2 && M_{114}^{(3)}=-2 \nonumber\\
&M_{122}^{(3)}=2 && M_{123}^{(3)}=1 && M_{124}^{(3)}=1 && M_{133}^{(3)}=-2 \nonumber\\
&M_{134}^{(3)}=1 && M_{223}^{(3)}=-2 && M_{234}^{(3)}=1 && M_{244}^{(3)}=-2 \nonumber\\
&M_{334}^{(3)}=2 && M_{344}^{(3)}=2. \label{eq:B444wp111}
\end{align}
In the Appendix we show that $\eta_{crit}$ is exactly $1/2$ for
this inequality. We have tried $m=5$ and $m=6$, but we have got no
improvement, so for three participants we could not find a Bell
inequality for which the critical efficiency goes below $1/2$.

We summarized critical detection efficiencies we found in this paper for the symmetric 3-qubit states in Table~\ref{table:bestdetsym}.

\begin{table}[t]
\vskip 0.2truecm
\begin{tabular}{c|c c}
\hline
$\text{settings}$&$\eta_{crit}$&$\text{equation}$\\
\hline
222&0.6&(\ref{eq:B222w})\\
223&0.6&(\ref{eq:Bell322})\\
333&0.51678&(\ref{eq:B333wp111})\\
444&0.5&(\ref{eq:B444wp111})\\
\hline
\end{tabular}
\caption{Table for critical detection efficiencies using symmetric 3-qubit states. The numbers in
brackets refer to the Bell inequalities. In the first column
$ijk$ refers to number of $i$,$j$, and $k$ settings for the
parties Alice, Bob, and Cecil. Detection efficiency thresholds are rounded up
to five digits.} \vskip 0.2truecm
\label{table:bestdetsym}
\end{table}

\vskip 0.5truecm

\section{Summary}

We have shown that the required detection efficiencies to
demonstrate a loophole-free Bell violation can be significantly
lowered if three parties are involved (instead of the usual
two-party scenario). Before, no practical three-party Bell tests
featuring efficiencies lower than $60\%$ were known to the best of our
knowledge. This value has been attained by Larsson and Semitocolos
in 2001 in a three-party two-setting Bell scenario \cite{LS}. We beat this
limit using a W state and three measurements per party. Moreover,
for 8 settings we reach the value of $50.13\%$. On the other hand,
using a coherent mixture of the W state with a product state
$|111\rangle$ allows us to obtain $\eta_{crit} = 50\%$ even with 4
settings. We conjecture that $\eta_{crit} = 50\%$ cannot be beaten in
either way.

It is left as an open question if one of our
inequalities could be generalized beyond three parties
similarly to the family of Bell inequalities by Larsson et al.
\cite{LS}.

\section{Acknowledgements}
We are grateful to Nicolas Brunner for useful discussions.
We acknowledge financial support from the Hungarian
National Research Fund OTKA (K111734), a J\'anos Bolyai Grant of
the Hungarian Academy of Sciences, SEFRI (COST action MP1006), and the
T\'AMOP-4.2.2.C-11/1/KONV-2012-0001 project. The project has also
been supported by the European Union, co-financed by the European
Social Fund.

\appendix
\section{Detailed calculation of critical detection efficiencies}\label{appendix}

In the Appendix we calculate the critical detector efficiencies
for the Bell inequalities given in the main text.  According to
Eqs.~(\ref{eq:violate},\ref{eq:2partcontr},\ref{eq:3partcontr}),
and taking into account the permutational symmetry of the states
considered, the matrix element of the effective Bell operator may
be written as:
\begin{align}
\langle\psi|\hat{\cal M}_\eta|\psi\rangle=\eta^2
\sum_{j=1}^{m}\sum_{i=j}^{m}M^{(2)}_{ij}\pi_{ij0}\langle\psi|\hat
A_i\otimes\hat B_j\otimes\hat I
|\psi\rangle+\nonumber\\
\eta^3 \sum_{k=1}^{m}\sum_{j=k}^{m}\sum_{i=j}^{m}M^{(3)}_{ijk}
\pi_{ijk}\langle\psi|\hat A_i\otimes\hat B_j\otimes\hat
C_k|\psi\rangle, \label{eq:app1}
\end{align}
where $\pi_{ijk}$ is the number of permutations of indices $i$,
$j$, and $k$, that is $\pi_{ijk}=6$, if all three are different,
$\pi_{ijk}=3$ if two indices agree, and $\pi_{ijk}=1$ if $i=j=k$.

The condition for the violation of the Bell inequality by
the results of the measurements performed on the $|W\rangle$ state is:
\begin{equation}
\langle W|\hat{\cal M}_\eta|W\rangle>0, \label{eq:appviolatew}
\end{equation}
and the values of the matrix elements necessary to evaluate
$\langle W|\hat{\cal M}_\eta|W\rangle$ for small measurement
angles are given by
Eqs.~(\ref{eq:wmatelemssmall1},\ref{eq:wmatelemssmall2}). It makes
the calculations simpler if we notice that these matrix elements
do not change if we reverse the signs of the measurement angles
concerned simultaneously. Also, if one of the measurement angles
is $\Phi$ and another one is $-\Phi$, then the three particles
matrix element will not depend on the third angle. These
statements are also true for the matrix elements shown in
Eq.~(\ref{eq:othermatelems}) in the limit of small angles, which
we will need when we consider the state defined by
Eq.~(\ref{eq:psistate}).

We have already shown that for the $m=3$ inequality given by
Eq.~(\ref{eq:B333w}) $\eta_{crit}=0.6$.

Now let us consider  the $m=4$ case given by
Eq.~(\ref{eq:B444w}). The measurement angles to be taken now are
$\Phi_i=\phi_ix$, with $\phi_1=-\phi_2=1$ and
$\phi_3=-\phi_4=\lambda$. By using
Eqs.~(\ref{eq:appviolatew},\ref{eq:app1},\ref{eq:wmatelemssmall1},\ref{eq:wmatelemssmall2}),
straightforward calculation leads us to:
\begin{equation}
-3-\lambda^4+\eta[6-3(1-2\lambda)^2]>0 \label{eq:appw444}
\end{equation}
for the condition of the quantum violation. Here we have
simplified the expression by a factor of $x^2\eta^2/(48\cdot 12)$.
At $\eta=\eta_{crit}$ the l.h.s.\ of the equation is zero,
therefore $\eta_{crit}=3(1+4\lambda-4\lambda^2)/(3+\lambda^4)$. It
has its minimum value if $\lambda$ satisfies
$2\lambda^5-3\lambda^4-\lambda^3-6\lambda+3=0$. The appropriate
root calculated numerically is $\lambda=0.466715$, which leads to
$\eta_{crit}=0.509036$.

For the the $m=6$ case shown in Eq.~(\ref{eq:B666w}) we can follow
the same steps as above. Now the measurement angles are given by
$\phi_1=-\phi_2=1$, $\phi_3=-\phi_4=\mu$ and $\phi_5=-\phi_6=\nu$.
With these angles we get for the condition of quantum violation,
after a simplification by a factor of $x^2\eta^2/(48\cdot 36)$:
\begin{align}
-3&(1+\mu^4+\nu^4)+\eta[6+6\mu^4+4\nu^4-3(1-2\mu)^2-\nonumber\\
3&(\mu^2-2\mu\nu)^2-3(\mu-\nu-\mu\nu)^2]>0. \label{eq:appw666}
\end{align}
Again, at $\eta=\eta_{crit}$ the l.h.s.\ of the equation is zero,
and we must choose the parameters $\mu$ and $\nu$ such that
$\eta_{crit}$ is minimal. We get three equations for the three
unknown values, and if we solve those equations numerically we get
$\mu=0.495815$, $\nu=0.295435$, and $\eta_{crit}=0.502417$.

For inequality with $m=8$ given by Eq.~(\ref{eq:B888w}) the expression
corresponding to Eq.~(\ref{eq:appw666}) is:
\begin{align}
-3&(1+\rho^4+\sigma^4+\tau^4)+\eta[6+6\rho^4+6\sigma^4+4\tau^4-\nonumber\\
3&(1-2\sigma)^2-3(\rho^2-2\rho\sigma)^2-3(\sigma^2-2\sigma\tau)^2-\nonumber\\
3&(\rho-\sigma-\rho\sigma)^2-3(\rho\sigma-\rho\tau-\sigma\tau)^2]>0. \label{eq:appw888}
\end{align}
Here we have followed the same steps as for $m=6$ taking measurement angles
$\phi_1=-\phi_2=1$, $\phi_3=-\phi_4=\rho$, $\phi_5=-\phi_6=\sigma$ and
$\phi_7=-\phi_7=\tau$. From the equation we get numerically $\eta_{crit}=0.501338$ with
$\rho=0.498442$, $\sigma=0.306395$ and $\tau=0.169989$.

Now let the state be the one shown in Eq~(\ref{eq:psistate}). From
Eq~(\ref{eq:qsmallxlimit}), if we choose the optimal mixing angle,
the condition for quantum violation is:
\begin{equation}
\langle W|\hat{\cal M}_\eta|W\rangle- \frac{\langle W|\hat{\cal
M}_\eta|111\rangle}{\langle 111|\hat{\cal M}_\eta|111\rangle}>0.
\label{eq:appviolatepsi}
\end{equation}
The matrix elements of the Bell operator may be calculated from
Eq~(\ref{eq:app1}), which is also valid if the state vectors are
different in the bra and the ket positions, provided both are
permutationally symmetric. The matrix elements of the two and
three particle operators appearing in the r.h.s.\ of the equation
are given in
Eqs.~(\ref{eq:wmatelemssmall1},\ref{eq:wmatelemssmall2},\ref{eq:othermatelems}).
We are concerned with the small angles limit.

First, let us take the $m=2$ inequality of Eq~(\ref{eq:B222w}).
With the choice of $\phi_1=0$ and $\phi_2=1$, we get $\langle
W|\hat{\cal M}_\eta|W\rangle=-3\eta^3x^4/48$, $\langle W|\hat{\cal
M}_\eta|111\rangle=-3\eta^3x^2/4\sqrt{3}$ and $\langle
111|\hat{\cal M}_\eta|111\rangle=-3\eta^2+2\eta^3$ for the matrix
elements of the effective Bell operator. By substituting these
values into Eq.~(\ref{eq:appviolatepsi}), and taking into account
that the l.h.s.\ of the equation is zero at $\eta=\eta_{crit}$, it
is easy to see that $\eta_{crit}=3/5=0.6$.

We may take the same steps for $m=3$. The inequality is shown by
Eq.~(\ref{eq:B333wp111}), and the measurement angles are given by
$\phi_1=0$, $\phi_2=1$ and $\phi_3=-1$. Then the matrix elements
of the effective Bell operator are $\langle W|\hat{\cal
M}_\eta|W\rangle=-6\eta^2x^4/48$, $\langle W|\hat{\cal
M}_\eta|111\rangle=-\sqrt{3}\eta^2x^2(2\eta-1/2)$ and $\langle
111|\hat{\cal M}_\eta|111\rangle=-\eta^2(12-5\eta)$. Then the
condition that the l.h.s.\ of Eq.~(\ref{eq:appviolatepsi}) is zero
at $\eta=\eta_{crit}$ leads to equation
$48\eta_{crit}^2-19\eta_{crit}-3=0$, whose appropriate root is
$\eta_{crit}=(19+\sqrt{937})/96\approx 0.516776$.

In the case of the $m=4$ inequality of Eq.~(\ref{eq:B444wp111})
the measurement angles are given by $\phi_1=1$, $\phi_2=-1$,
$\phi_3=\lambda$ and $\phi_4=-\lambda$. From these it follows that
$\langle W|\hat{\cal M}_\eta|W\rangle=
\eta^2x^4[-1-\lambda^4+\eta(1+4\lambda-8\lambda^2-4\lambda^3+\lambda^4)]/4$,
$\langle W|\hat{\cal
M}_\eta|111\rangle=-\sqrt{3}\eta^2x^2(1+\lambda^2)(1-3\eta)$ and
$\langle 111|\hat{\cal M}_\eta|111\rangle=-24\eta^2(1-\eta)$. If
we substitute these values into Eq.~(\ref{eq:appviolatepsi}), we
can get:
\begin{equation}
\frac{\eta^2}{8(1-\eta)}\bigg [r\big (\eta-\frac{1}{2}\big
)^2+p\big (\eta-\frac{1}{2}\big )-q\bigg ]>0,
\label{eq:app444wp111}
\end{equation}
where
\begin{align}
r&\equiv 4\lambda^4+8\lambda^3+34\lambda^2-8\lambda+7\nonumber\\
&=5\lambda^4+2(\lambda+1)^4+6\lambda^2+4(2\lambda-1)^2+1>0\nonumber\\
p&\equiv 5\lambda^4+6\lambda^2+5>0\nonumber\\
q&\equiv\frac{(\lambda^2+4\lambda-1)^2}{4}\geq 0.
\label{eq:app444wp111a}
\end{align}
In Eq.~(\ref{eq:app444wp111}) the prefactor is positive for
$0<\eta<1$. As $r$ is strictly positive, the inequality is
satisfied above the upper root of the second order expression.
Below that the expression is negative for all $\eta\geq 0$, as one
can easily see. Therefore, we get the critical efficiency as
$(\eta_{crit}-1/2)=(\sqrt{p^2+4rq}-p)/2r$. As $r>0$, $p>0$ and
$q\geq 0$, the smallest possible value the r.h.s.\ may take is
zero, when we choose $\lambda$ such that $q=0$, that is
$\lambda=-2\pm\sqrt{5}$. With this optimal choice
$\eta_{crit}=1/2$.

\end{document}